\documentclass[11pt,twoside]{article}

\usepackage{asp2006}
\usepackage{epsf}
\usepackage{psfig}
\usepackage{lscape}

\begin{document}
\title{The Algol type eclipsing binary X Tri:\\
BVRI modelling and O-C diagram analysis}   
\author{Liakos, A.$^{1}$, Zasche, P.$^{2,3}$ and Niarchos, P.$^{1}$}   
\affil{$^{1}$Department of Astrophysics, Astronomy and Mechanics,
National and Kapodistrian University of Athens, GR 157 84
Zografos, Athens, Greece\\
$^{2}$Astronomical Institute, Faculty of Mathematics and Physics, Charles University Prague, CZ-180 00 Praha 8, V Hole\v{s}ovi\v{c}k\'{a}ch 2, Czech Republic\\
$^{3}$Instituto de Astronom\'{i}a, Universidad Nacional Aut\'{o}noma de M\'{e}xico, A.P. 70-264, M\'{e}xico, DF 04510, Mexico }    

\begin{abstract} 
CCD photometric observations of the Algol-type eclipsing binary X
Tri have been obtained. The light curves are analyzed with the
Wilson-Devinney (WD) code and new geometric and photometric
elements are derived. A new O-C analysis of the system, based on
the most reliable timings of minima found in the literature, is
presented and apparent period changes are discussed with respect
to possible and multiple Light-Time Effect (LITE) in the system.
Moreover, the results for the existence of additional bodies
around the eclipsing pair, derived from the period study, are
compared with those for extra luminosity, derived from the  light
curve analysis.

\end{abstract}



\section{Observations and analyses}

The system was observed during 5 nights from October 2008 to
January 2009 at the Athens University Observatory, using a 40-cm
Cassegrain telescope equipped with the CCD camera ST-8XMEI and B,
V, R, I Bessell filters.

The light curves (hereafter LCs) were analysed with the $PHOEBE~
0.29d$ software \citep{PZ05} which uses the 2003 version of the WD
code \citep{WD71,W79}. Due to the lack of spectroscopic mass ratio
of the system, the q-search method was applied in Mode 2, 4 and 5
in order to find the most probable value of the (photometric) mass
ratio (q).

The least squares method with statistical weights has been used
for the analysis of the O-C diagram. The current O-C diagram of X
Tri includes 572 times of minima taken from the literature. The
following ephemeris: $Min.I=HJD~2422722.285 + 0.9715341^{d} \times
E$ \citep{K01} was used as the initial one for the O-C analysis of
the compiled times of minima.

\section{Discussion and Conclusions}

The results of the LCs solution (see figure 1 and table 1) show
that X Tri is a semi-detached system with the secondary star
filling its Roche Lobe. The periodic variations of the orbital
period of the system could be explained by adopting the existence
of three additional components, which were found to have minimal
masses 0.18, 0.24 and 0.22, respectively (see figure 1 and table
2). An extra light contribution to the luminosity of the EB was
taken into account in the LCs solution but it was found to be less
than 1\%. Such a small extra luminosity could be explained by the
small values of the minimal masses of the possible additional
components found. The steady decrease rate of its period is
probably due to angular momentum loss, since the direction of the
flow (from the more massive to the less massive component),
revealed from the O-C diagram analysis, comes in disagreement with
the one derived from the LCs analysis.

\begin{table}

\caption{The parameters of X Tri derived from the LCs solutions}
\label{tab1} \scalebox{0.58}{
\begin{tabular}{ccccccc}

\hline
\textbf{Parameter}          & \textbf{value} & \textbf{Parameter} & \multicolumn{4}{c}{\textbf{value}}     \\
\hline
$q~(m_{2}/m_{1})$           &    0.599~(2)   &                    &    B    &    V    &    R    &    I     \\
$i$~(deg)                   &    87.9~(1)    &   $x_{1}$$^{***}$  &  0.551  &  0.478  &  0.402  &  0.322   \\
$T_1$$^{**}$(K), $T_2$ (K)  & 8600, 5188~(4) &   $x_{2}$$^{***}$  &  0.835  &  0.692  &  0.597  &  0.503   \\
$A_1$$^*$, $A_2$$^*$        &     1, 0.5     &  $L_{1}/L_{T}$     &0.893~(2)&0.839~(2)&0.795~(2)&0.739~(1) \\
$g_1$$^*$, $g_2$$^*$        &     1, 0.32    &  $L_{2}/L_{T}$     &0.107~(1)&0.160~(2)&0.201~(2)&0.246~(3) \\
$\Omega_{1}$, $\Omega_{2}$  & 4.27~(1), 3.06 &  $L_{3}/L_{T}$     &0.000~(1)&0.000~(1)&0.004~(1)&0.015~(2) \\
$\chi^{2}$                  &     1.278      &                    &                                        \\
\hline \textit{$^*$assumed},&\textit{$^{**}$ \citet{G83},}
&\textit{$^{***}$\citet{VH93}},& \textit{$L_{T} =
L_{1}+L_{2}+L_{3}$}

\end{tabular}}

\end{table}

\begin{table}

\caption{The results of the O-C diagram analysis for X Tri}
\label{tab2} \scalebox{0.52}{

\begin{tabular}{cccccc}

\hline
\textbf{Parameters of the EB} & \textbf{value}  & \textbf{Parameters of the LITEs} &               \multicolumn{3}{c}{\textbf{value}}                \\
\hline
$M_1^{*}+M_2~(M_\odot)$       &  2.1 + 1.26     &                                  &      $3^{rd} body$    &     $4^{th} body$    &    $5^{th} body$ \\
$Min. I$~(HJD)                & 2442502.731~(2) &           $P$~(yrs)              &        36.9~(5)       &        22.4~(3)      &       16.8~(4)   \\
$P$~(days)                    & 0.9715318~(2)   &   $T_0$~(HJD),~$\omega$~(deg)    &2452916~(373), 220~(98)&2455069~(335), 34~(13)&     --, --       \\
$c_{2}~(\times 10^{-10})$     &  -2.0308~(2)    &        $A$~(days),~$e$           &  0.0052~(3), 0.2~(2)  &  0.0040~(4), 0.5~(1) &   0.003~(2), 0.0 \\
$\dot{P}~(\times 10^{-10})$   &  -1.5269~(2)    &      $M_{min}~(M_\odot)$         &        0.18~(1)       &       0.24~(1)       &      0.22~(1)    \\
\hline
\textit{$^*$assumed}

\end{tabular}}

\end{table}

\begin{figure}[ht!]
\plottwo{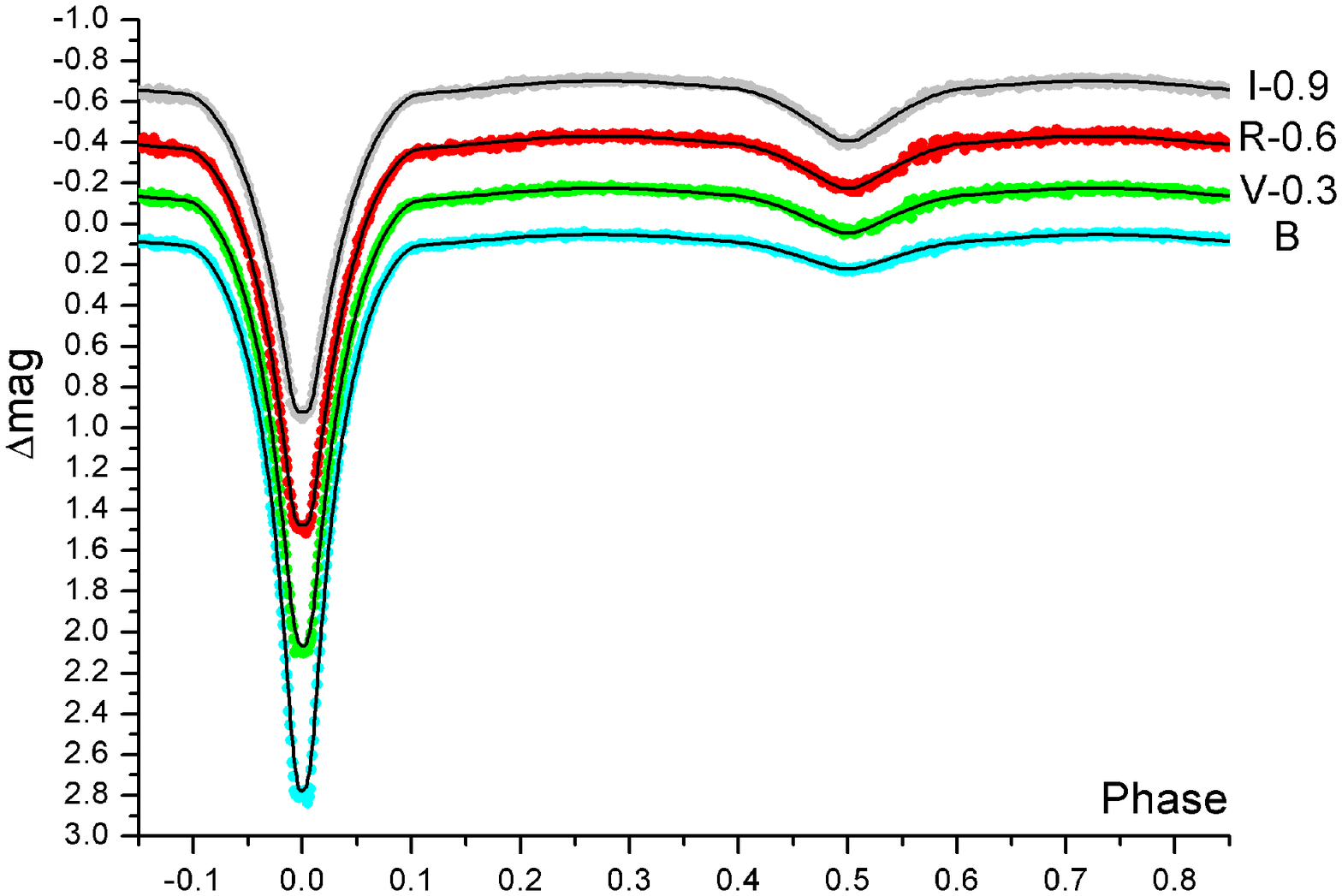}{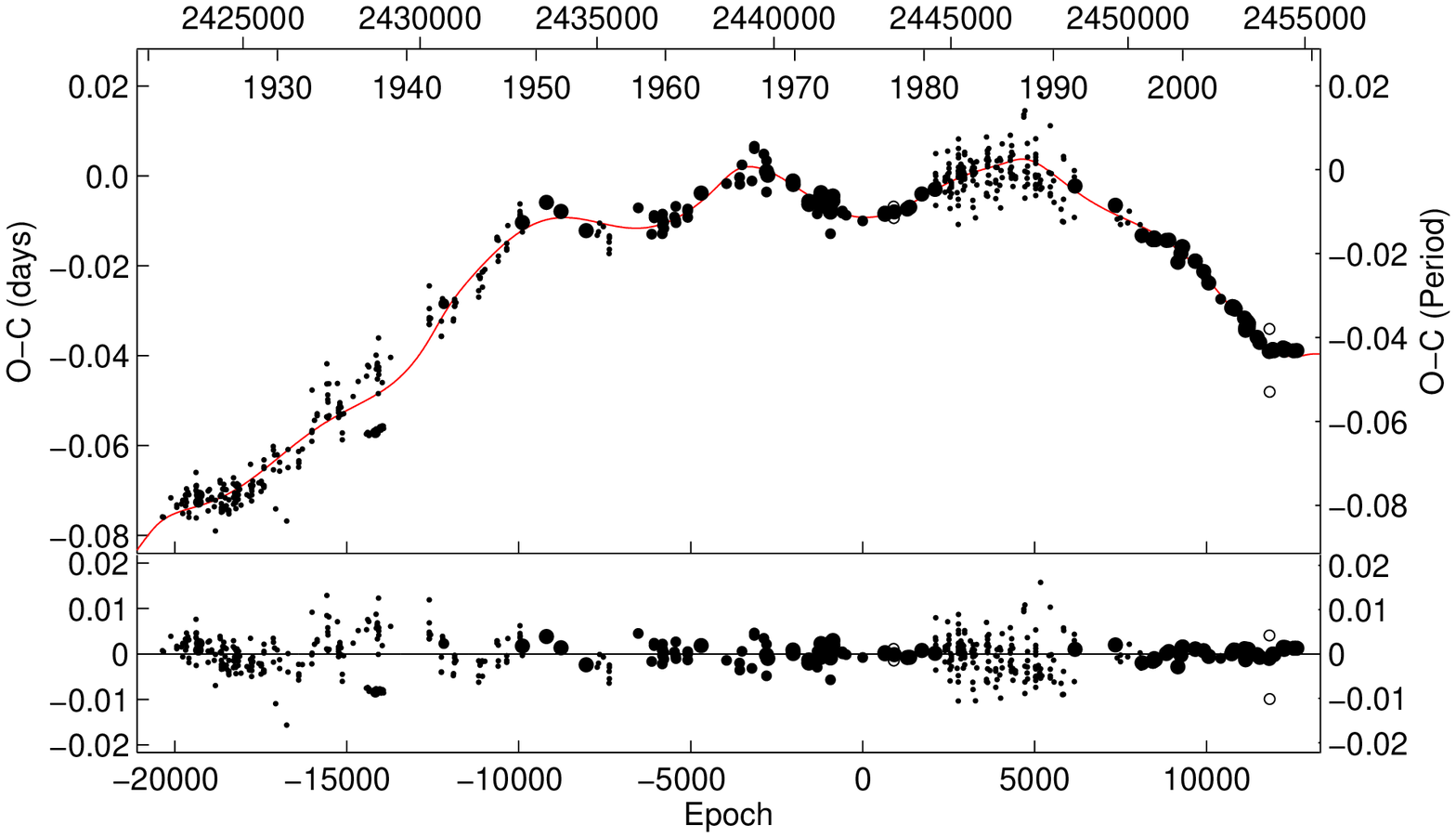}

\caption{Left panel: The synthetic LCs (black solid lines) along
with the observed ones (colored points). Right panel: The
multiperiodic fitting (red solid line) on the O-C points (black
points) (upper part) and the O-C residuals (lower part).}
\end{figure}



\end{document}